\documentstyle[aps,prd,graphicx,axodraw]{revtex}
\begin{document}
\hfill{hep-ph/0211272}\par
\hfill{DPNU-02-35}\par
\vskip 0.3cm
\centerline{\large \bf Threshold resummation for nonleptonic
$B$ meson decays}
\vskip 0.3cm
\centerline{Hsiang-nan Li$^{1}$\footnote{E-mail: hnli@phys.sinica.edu.tw} and
Kazumasa Ukai$^2$\footnote{E-mail: ukai@eken.phys.nagoya-u.ac.jp}}\par
\centerline{$^1$Institute of Physics, Academia Sinica,
Taipei, Taiwan 115, Republic of China}
\centerline{$^1$Department of Physics, National Cheng-Kung University, 
Tainan, Taiwan 701, Republic of China}\par
\centerline{$^2$Department of Physics, Nagoya University, 
Nagoya, 464-8602, Japan}\par
\vskip 0.5cm

\vskip 0.5cm

\centerline{\bf Abstract}
\vskip 0.3cm

We investigate the double logarithmic corrections $\alpha_s\ln^2 x$, 
$x$ being a parton momentum fraction, in two-body nonleptonic $B$ meson 
decays in collinear factorization theorem of perturbative QCD (PQCD).
It is found that these corrections are universal for
factorizable amplitudes, {\it i.e.}, at the leading power, and that 
their threshold resummation smears the end-point singularities from 
$x\to 0$. The double logarithmic corrections, depending on the topologies
of nonfactorizable amplitudes at the subleading power, are
negligible due to the color-transparency argument and to the 
overlap of three meson distribution amplitudes. We show that the PQCD 
approach to two-body nonleptonic $B$ meson decays respects the
factorization assumption in the heavy-quark limit.

\vskip 1.0cm

\section{INTRODUCTION}

It has been shown that fixed-order evaluation of $B$ meson semileptonic
decay amplitudes in the framework of collinear factorization theorem
suffers the end-point singularities from a parton momentum fraction 
$x\to 0$ \cite{SHB,BD,BF}. On the other hand, the double logarithms 
$\alpha_s\ln^2 x$ appearing in higher-order corrections to these
decays have been observed \cite{ASY,KPY,L3,DS,LPW}. We argue that when 
the end-point region is important, $\alpha_s\ln^2 x$ can not be treated 
as a small expansion parameter, and should be summed to all orders. A 
systematic treatment of these logarithms has been proposed by grouping 
them into a quark jet function, whose dependence on $x$ is governed by 
an evolution equation \cite{L3}. A Sudakov form factor, obtained by 
solving the evolution equation, decreases fast enough at the end point. 
The above procedure is referred to as threshold resummation, whose
technical details can be found in \cite{S0,CT,L2}. It turns out
that in a self-consistent analysis, where the original factorization 
formulas \cite{BL,ER,CZS,L4} are convoluted with the Sudakov factor, the 
end-point singularities do not exist \cite{TLS}. Therefore, it is not 
necessary to introduce arbitrary infrared cutoffs for momentum fractions 
$x$ even in the collinear factorization theorem, such as the QCD 
factorization(QCDF) approach \cite{BBNS} to exclusive $B$ meson decays.
The same observation has been made recently in the framework of
soft-collinear effective theory \cite{CPS}.

In this letter we shall examine the double logarithmic corrections to 
all the topologies of two-body nonleptonic $B$ meson decay 
amplitudes, including charmless and charmful modes. The topologies
contain both emission and annihilation, which are further divided into 
factorizable and nonfactorizable types. Our results for the charmless
and charmful decays are the same, and summarized as follows. The double 
logarithmic corrections are crucial for factorizable (regardless of
charmless or charmful, emission or annihilation) contributions due to
the presence of the potential linear end-point singularities. As to
nonfactorizable contributions, the double logarithmic corrections exist,
and carry a color factor different from that in the factorizable cases. 
However, they are less crucial for charmless decays, since the 
end-point singularities are at most logarithmic: there exists soft 
cancellation between a pair of nonfactorizable emission diagrams near 
the end point due to the color-transparency argument \cite{Bj}. For
charmful decays, the hierarchy that the $b$ quark mass is much larger
than the $c$ quark mass is the necessary condition for perturbative QCD
(PQCD) to be applicable \cite{TLS2}. Under this hierarchy, the soft
cancellation between a pair of nonfactorizable emission amplitudes holds
approximately. The factorization formulas for nonfactorizable (charmless
and charmful) annihilation amplitudes involve the overlap integrals of
three meson distribution amplitudes, such that the end-point
singularities become milder.

In Sec.~II we identify the double logarithms involved in the various 
topologies of decay amplitudes. The numerical analysis is performed
in Sec.~III. We show that threshold resummation is not required for
nonfactorizable amplitudes, since the logarithmic singularities 
can be easily smeared by the Sudakov effect from $k_T$ resummation 
\cite{LY1,KLS,LUY}. Note that threshold resummation deals with the double 
logarithmic corrections to hard amplitudes, while $k_T$ resummation deals
with those to hadron wave functions. Hence, the former may not be 
universal as stated above, but the latter must be, since wave functions 
are. Both effects are necessary, especially when the end-point
singularity is linear. Considering either of them, though smears the
singularity \cite{L3,CPS}, does not suppress the soft contribution
strongly enough \cite{TLS}. The results derived in this work provide a
useful reference for the study of two-body nonleptonic $B$ meson decays. 
At last, we explicitly demonstrate that the PQCD formalism for
two-body nonleptonic $B$ meson decays reduces to the factorization
assumption \cite{BSW} in the heavy-quark limit. Section IV is the
conclusion.

\section{DOUBLE LOGARITHMS}

We take the $B\to K\pi$ decays as an example. The momenta of the $B$
meson, kaon and pion, and of their valence quarks are chosen as
\begin{eqnarray}
P_1&=&(M_B/\sqrt{2})(1,1,{\bf 0}_T)\;,\;\;\;\;
k_1=(M_B/\sqrt{2})(x_1,0,{\bf 0}_T)\;,
\nonumber\\
P_2&=&(M_B/\sqrt{2})(1,0,{\bf 0}_T)\;,\;\;\;\;
k_2=x_2P_2\;,
\nonumber\\
P_3&=&(M_B/\sqrt{2})(0,1,{\bf 0}_T)\;,\;\;\;\;
k_3=x_3P_3\;,
\label{pp}
\end{eqnarray}
respectively, $x_{1,2,3}$ being the momentum fractions. The parton
momentum $k_1$ ($k_3$) is associated with the spectator quark in the
$B$ meson (pion). The momentum $k_2$ is associated with the light quark 
(not the $s$ quark) in the kaon. The mass difference between the $B$
meson and the $b$ quark, $\bar\Lambda=M_B-m_b$, is treated as a
small scale in the following analysis. 
To concentrate on the double logarithm $\alpha_s\ln^2 x$, we adopt
the collinear factorization theorem, and do not
include the transverse momenta $k_T$ of the valence quarks.

If the end-point region is defined by $x\to 0$, the double logarithm, 
being of the collinear origin, is produced through the loop integral
in the covariant gauge,
\begin{eqnarray}
J^{(1)}&=&-ig^2C\int\frac{d^4 l}{(2\pi)^4}
\frac{n_\pm \cdot n}{n_\pm\cdot(l-xP)n\cdot l l^2}\;,
\nonumber\\
&=&-\frac{\alpha_s}{4\pi}C\ln^2 x+\cdots\;,
\label{j2c}
\end{eqnarray}
where $C$ is a color factor depending on the topology, the dimensionless
vectors $n_+=(1,0,{\bf 0}_T)$ and $n_-=(0,1,{\bf 0}_T)$ on the light 
cone, $n$ an arbitrary dimensionless vector, and $xP$ the 
fractional momentum carried by a parton. The Feynman rules
$n_\pm^\alpha/n_\pm\cdot(l-xP)$ and $n_\alpha/n\cdot l$ arise from the 
eikonal approximation of quark propagators. Their appearance will be
explained below. A double logarithm is associated with a system
containing two scales \cite{L2}: one is of $O(\Lambda_{\rm QCD})$ and
another is a large momentum transfer. With this principle, it is easy
to identify the source of double logarithms in the different topologies 
of diagrams. 

Our analysis of the double logarithmic corrections applies to
charmful $B$ meson decays, such as $B\to D\pi$. It has been argued that
the PQCD formalism for these decays holds under the hierarchy \cite{TLS2},
\begin{eqnarray}
M_B\gg M_D\gg \bar\Lambda\;,
\label{hie}
\end{eqnarray}
with $M_D$ being the $D$ meson mass.
The relation $M_B\gg M_{D}$ justifies the perturbative calculation
of the $B\to D$ transition at large recoil and the definition of
light-cone $D$ meson wave functions. The relation 
$M_{D}\gg\bar\Lambda$ justifies the power expansion in the 
parameters $\bar\Lambda/M_{D}$ and $\bar\Lambda/M_{B}$. Therefore,
the $B\to D$ hard amplitudes are basically the same as the
$B\to\pi$ ones at the leading power.
It has been shown that the charm mass effect
appears through the ratio $(M_D/M_B)^2\sim 0.1$,
which is indeed negligible.

\subsection{Factorizable Diagrams}

The lowest-order factorizable and nonfactorizable diagrams for the 
$B\to K\pi$ decays are displayed in Fig.~\ref{fig:leading_Kpi}. We start
with the factorizable emission diagram in Fig.~\ref{fig:leading_Kpi}(a), 
which gives a hard amplitude proportional to $1/(x_1x_3^2)$. Obviously, 
this amplitude, as convoluted with the pion distribution amplitude 
$\phi_\pi(x_3)$, leads to a logarithmic divergence for 
$\phi_\pi(x_3)\sim x_3$ (twist-2) and to a linear divergence for 
$\phi_\pi(x_3)\sim 1$ (two-parton twist-3) at small $x_3$.
As $x_3\to 0$ (to be precise, $x_3\sim O(\bar\Lambda/M_B)$), the 
internal $b$ quark, carrying the momentum $P_1-x_3P_3$, approaches the 
mass shell. This is the origin of the end-point singularity. The double
logarithms in radiative corrections to Fig.~\ref{fig:leading_Kpi}(a)
then become important. Part of the 
$O(\alpha_s)$ corrections are shown in Fig.~\ref{fig:higher_a}. 
The discussion of these diagrams is the same as that for the
$B\to\pi$ form factors \cite{L3}, and the results are summarized
below. 

Figure \ref{fig:higher_a}(a) generates the double logarithm
$\alpha_s\ln^2 x_3$ through Eq.~(\ref{j2c}) with the color factor
$C=C_F=4/3$. In this case the $b$
quark propagator gives $n_+^\alpha/n_+\cdot(l-x_3P_3)$ under the
eikonal approximation in the collinear region with the loop momentum
$l$ being parallel to $P_3$. The light quark propagator gives
$n_\alpha/n\cdot l$ with $n$ being almost in the direction of $P_3$. 
A small plus component has been included into $n$ to regularize the
infrared pole of Eq.~(\ref{j2c}). The self-energy correction in 
Fig.~\ref{fig:higher_a}(b), and the vertex corrections in 
Figs.~\ref{fig:higher_a}(c) and \ref{fig:higher_a}(d) produce only the
single logarithm $\alpha_s\ln x_3$, since these diagrams do not involve
a large momentum transfer. The double logarithms are produced from
the vertex corrections in Figs.~\ref{fig:higher_a}(e) and
\ref{fig:higher_a}(f) with the gluons attaching the kaon. However, they
cancel each other due to the opposite signs in the eikonal approximations
for a quark and for an anti-quark. Note that 
Fig.~\ref{fig:higher_a}(d) contains the double logarithm 
$\alpha_s\ln^2 x_1$ as shown in \cite{ASY}. However, this double 
logarithm is not very relevant, because the convolution in $x_1$,
with the corresponding hard amplitude 
$1/(x_1x_3^2)$, diverges at most logarithmically.

Figure~\ref{fig:leading_Kpi}(b) gives an amplitude proportional to 
$1/(x_1^2 x_3)$. The factorization formula diverges linearly, if using
the $B$ meson distribution amplitudes derived in \cite{GN,KQT}. This
end-point singularity indicates that collinear enhancements arise 
from the higher-order corrections associated with the internal light 
quark as $x_1\sim O(\bar\Lambda^2/M_B^2)$. The analysis is similar to 
that for Fig.~\ref{fig:leading_Kpi}(a): only the diagram, in which the
radiative gluon attaches the $b$ quark and the internal quark carrying
the momentum $P_3-k_1$, generates the double logarithm
$\alpha_s\ln^2 x_1$. In this case the internal light quark propagator
gives $n_-^\alpha/n_-\cdot(l-k_1)$ in Eq.~(\ref{j2c}), $n$ stands
for the $b$ quark velocity $v=P_1/M_B$, and the color factor is
$C=C_F$.

The above study of double logarithms applies to the factorizable 
annihilation diagrams straightforwardly. The lowest-order diagram in 
Fig.~\ref{fig:leading_Kpi}(c) gives a hard amplitude proportional to
$1/(x_2{\bar x}_3^2)$, ${\bar x}_3\equiv 1-x_3$. Hence, the end-point 
region is defined by ${\bar x}_3\to 0$, where additional collinear 
divergences are associated with the internal $s$ quark carrying the
momentum $P_2+{\bar x}_3P_3$. The loop correction to the weak decay
vertex, in which the radiative gluon attaches the internal $s$ quark and
the light spectator quark flowing into
the pion, produces the double logarithm $\alpha_s\ln^2 {\bar x}_3$. 
The internal $s$ quark propagator leads to 
$n_+^\alpha/n_+\cdot(l+{\bar x}_3P_3)$ in Eq.~(\ref{j2c}), the vector $n$ 
denotes the direction almost parallel
to $P_3$, and the color factor is $C=C_F$.
The double logarithms in the pair of diagrams with the radiative gluon 
attaching the $b$ quark and the soft spectator quark cancel, since a
$B$ meson, dominated by soft dynamics, remains color transparent for 
collinear gluons. The discussion of the double logarithmic corrections
from the other next-to-leading-order diagrams is similar: they generate 
only the single logarithms.

For Fig.~\ref{fig:leading_Kpi}(d) with the internal quark carrying the
momentum $P_3+x_2P_2$, the end-point region corresponds to $x_2\to 0$.
The loop correction to the weak decay vertex, in which the radiative
gluon attaches the $s$ quark and the internal quark, produces the double
logarithm $\alpha_s\ln^2 x_2$ with the color factor
$C=C_F$. The internal light quark 
propagator gives $n_-^\alpha/n_-\cdot(l+x_2P_2)$ in Eq.~(\ref{j2c}), and 
the vector $n$ denotes the direction almost parallel to $P_2$. The 
discussion for the other diagrams is similar to that for 
Fig.~\ref{fig:leading_Kpi}(c).
It is then concluded that the double logarithms $\alpha_s\ln^2 x$ are 
universal for the factorizable amplitudes, and 
independent of the color structures of the four-fermion operators.

\subsection{Nonfactorizable Diagrams}

In Fig.~\ref{fig:leading_Kpi}(e) the
internal $s$ quark carries the momentum ${\bar x}_2P_2-k_1+x_3P_3$,
${\bar x}_2\equiv 1-x_2$. There are three kinematic regions, in which
the $s$ quark approaches the mass shell, and additional infrared 
divergences appear:
\begin{eqnarray}
&(1) &\;\;x_1\sim {\bar x}_2\sim x_3\sim O(\bar\Lambda/M_B)\;,
\nonumber\\
&(2) &\;\;x_1\sim x_3\sim O(\bar\Lambda/M_B)\;,\;\;\;\;
{\bar x}_2\sim O(1)\;,
\nonumber\\
&(3) &\;\;x_1\sim {\bar x}_2\sim O(\bar\Lambda/M_B)\;,\;\;\;\;
x_3\sim O(1)\;.
\end{eqnarray}
In the first configuration no double logarithm is generated due to
the lack of a large scale. Because the hard gluon propagator is
proportional to $1/(x_1x_3)$, the contribution to the hard amplitude from
the third configuration is power-suppressed compared to that from the 
second one. Therefore, the end-point region is defined by the second 
configuration, in which additional collinear divergences from the loop 
momentum parallel to $P_2$ are generated. 

The vertex corrections in Figs.~\ref{fig:nonfact_cort}(a) and
\ref{fig:nonfact_cort}(b) contain the double
logarithms $\alpha_s\ln^2 x_3$. The internal $s$ quark propagator leads 
to $n_+^\alpha/n_+\cdot(l+x_3P_3)$ in Eq.~(\ref{j2c}), and the vector
$n$ stands for the $b$ quark velocity $v$ in the former, and for the 
direction almost along $P_3$ in the latter. If there are two 
separate color flows between the kaon and the $B\to\pi$ transition form
factor, the two double logarithms cancel each other, since they have 
the same color factor, but are opposite in sign. If 
there is only a single color flow, the two double logarithms possess
different color factors and add together. The result is proportional
to the net color factor $C=C_F+1/(2N_c)=N_c/2$, $N_c=3$ being the number
of colors. The other diagrams give only the single logarithms.
In Fig.~\ref{fig:leading_Kpi}(f), the internal light quark carries the
momentum $x_2P_2-k_1+x_3P_3$. Hence, the end-point region is defined by
the configuration,
\begin{eqnarray}
x_1\sim x_3\sim O(\bar\Lambda/M_B)\;,\;\;\;\;x_2\sim O(1)\;.
\end{eqnarray}
Additional collinear divergences and the double logarithm
$\alpha_s\ln^2 x_3$ with the color factor $C=N_c/2$ occur, when the 
diagram involves only a single color flow. This result is the same as 
for Fig.~\ref{fig:leading_Kpi}(e).

We emphasize that the factorization formula for the nonfactorizable 
annihilation diagrams suffers an ambiguity in defining a light-cone 
$B$ meson distribution amplitude. In the factorizable emission diagrams 
only the kinematics of the pion is relevant, whose momentum specifies a 
unique light-cone direction. The hard amplitudes then contain only the
variable $k_1^+$ through the inner product $k_1\cdot P_3$. An unambiguous
light-cone $B$ meson distribution amplitude $\phi_B(x_1)$, depending
on $x_1=k_1^+/P_1^+$, can be defined. For the nonfactorizable emission
diagrams, the kaon kinematics is also relevant, and the $k_1^-$
dependence appears in the hard amplitudes through $k_1\cdot P_2$.
Fortunately, $k_1^-$, less essential than $k_1^+$ \cite{BBNS}, is
negligible, and $\phi_B(x_1)$ can be defined. In the factorizable
annihilation diagrams the $B$ meson dynamics decouples. In the
nonfactorizable annihilation diagrams, differing from
the other topologies, the outgoing kaon and pion specify two different
light-cone directions (one in the plus direction and another in the minus
direction). Hence, both the components $k_1^+$ and $k_1^-$ are important
in the hard amplitudes. However, we argue that the current leading-power 
formalism lives with such an ambiguity present at the subleading level.
For the discussion below, we simply assume that $k_1$ lies in the
plus direction as indicated in Eq.~(\ref{pp}). 

Since a $b$ quark is heavy, there is no soft cancellation between
Figs.~\ref{fig:leading_Kpi}(g) and \ref{fig:leading_Kpi}(h). However,
the factorization formulas are overlap integrals of three meson
distribution amplitudes, so that the end-point singularities are at
most logarithmic. In Fig.~\ref{fig:leading_Kpi}(g), the internal $b$ 
quark carries the momentum $P_1-k_1-x_2P_2-\bar{x}_3P_3$. To have this 
internal quark on the mass shell, we demand
\begin{eqnarray}
x_1\sim x_2 \sim \bar{x}_3\sim O(\bar\Lambda/M_B)\;.
\end{eqnarray}
In this case the $B$ meson momentum $P_1$ provides the large scale. 
The two diagrams, in which the radiative gluon emitted from the $b$
quark attaches the $s$ quark and the spectator quark in the pion,
contribute constructively. The double logarithm appears as
$\alpha_s\ln^2(x_2 +{\bar x}_3)$ with the color factor $C=N_c/2$. Because
$x_2$ and ${\bar x}_3$ must be small simultaneously in order to have
a large double logarithm, the phase space is restricted, and the
resummation effect is expected to be negligible.

The virtual spectator quark in Fig.~\ref{fig:leading_Kpi} (h) carries
the momentum $k_1-x_2P_2-{\bar x}_3P_3$. A similar investigation shows 
that the dominant region for the corresponding factorization formula is 
the one with $x_2\sim O(\bar\Lambda/M_B)$ and ${\bar x}_3\sim O(1)$ at 
the leading twist. This is consistent with our assumption of considering 
only the plus component $k_1^+$, which is selected by the pion
momentum $P_3$. Therefore, the end-point region for 
Fig.~\ref{fig:leading_Kpi}(h) is defined by 
\begin{eqnarray}
x_1\sim x_2\sim O(\bar\Lambda/M_B)\;, \;\;\;\;
\bar{x}_3\sim O(1)\;. 
\end{eqnarray}
In this region the diagram with the radiative gluon attaching the
internal spectator quark and the $s$ quark 
generates the double logarithm $\alpha_s\ln^2 x_2$
with the color factor $C=1/(2 N_c)$. 
The diagram with the radiative gluon attaching the
internal spectator quark and the spectator quark in the pion
produces only the single logarithm. Since the double logarithm has a
smaller color factor $1/(2N_c)$, the resummation effect can be dropped.
It is obvious from the above discussion that the double logarithms for
the nonfactorizable diagrams are not universal.

\section{NUMERICAL ANALYSIS}

It has been stated that the double logarithmic corrections 
to the nonfactorizable amplitudes are less essential.
Below we shall proceed a quantitative study.
As shown in \cite{L3}, the threshold resummation of the double logarithms 
introduces a Sudakov factor $S_t(x)$ into the PQCD factorization formulas 
near the end points,
\begin{eqnarray}
S_t(x)=\int_{a-i\infty}^{a+i\infty}\frac{dN}{2\pi i}
\frac{J(N)}{N}(1-x)^{-N}\;,
\label{mj}
\end{eqnarray}
where $a$ is an arbitrary real constant larger than all the real
parts of the poles involved in the integrand. The function $J(N)$ is
given, to the leading logarithm, by
\begin{eqnarray}
J(N)=\exp\left[\frac{1}{2}\int_0^1dz\frac{1-z^{N-1}}{1-z}
\int_{(1-z)}^{(1-z)^2}
\frac{d\lambda}{\lambda}\gamma_K(\alpha_s(\sqrt{\lambda M_B^2/2}))\right]\;,
\label{exa}
\end{eqnarray}
with the anomalous dimension $\gamma_K=C \alpha_s/\pi$.
After deriving the Sudakov factors from the threshold resummation and from
$k_T$ resummation for two-body nonleptonic $B$ meson decays, we perform
the numerical analysis. For this purpose,
we propose the parametrization for $S_t(x)$ \cite{TLS,LL2},
\begin{equation}
S_t(x)=\frac{2^{1+2 c}\Gamma(\frac{3}{2}+c)}{\sqrt{\pi} \Gamma(1+c)}
[x(1-x)]^c\;,
\label{st}
\end{equation}
where the parameter $c$ is determined by fitting the Mellin 
transformation of the above expression to $J(N)$ in Eq.~(\ref{exa}).
The fitted parameters are about $c=0.27$ for the factorizable diagrams, 
$0.31$ for the nonfactorizable emission diagrams, and 
$0.03$ for the nonfactorizable annihilation diagram in 
Fig.~\ref{fig:leading_Kpi}(h). For Fig.~\ref{fig:leading_Kpi}(g),
the Sudakov factor associated with the double logarithm
$\alpha_s\ln^2(x_2 +{\bar x}_3)$ can not be parametrized by
Eq.~(\ref{st}). We simply consider Fig.~\ref{fig:leading_Kpi}(h)
as an representative example to demonstrate the tiny 
resummation effect for the nonfactorizable annihilation diagrams.

The PQCD factorization formulas for the various topologies of decay
amplitudes are referred to \cite{LUY,CKL}. The double logarithms and
their threshold resummation effects on the diagrams in
Fig.~\ref{fig:leading_Kpi} are summarized in
Table~\ref{tb:num_eff}. It is observed that the Sudakov factor
decreases the factorizable emission and annihilation amplitudes by 
at least 40\%. The resummation effect can also be as large as 40\% for the 
imaginary part of each nonfactorizable emission diagram, but reduces to 
about 20\% for their sum. This confirms our argument that the threshold 
resummation is less important due to the soft cancellation at the end 
points. There is almost no effect on the nonfactorizable annihilation 
diagram as expected. Since the corresponding hard amplitudes oscillate 
between positive and negative values, the convolution with $S_t$ may
not always decrease.

\begin{table}[htbp]
 \begin{tabular}[tb]{|c|c|c|c|}
  & double logs.  & amplitudes without $S_t$ & amplitudes with $S_t$ \\
  \hline
  \hline
  1(a) & $-(\alpha_s/4\pi)C_F \ln^2 x_3$ & $1190.9$ & $679.6$ \\
  1(b) & $-(\alpha_s/4\pi)C_F \ln^2 x_1$ & $394.8$ & $306.7$ \\
  \hline
  1(a)+1(b) & & $1585.8$ & $986.4$ \\
  \hline
  \hline
  1(c) & $-(\alpha_s/4\pi)C_F \ln^2 {\bar x}_3$ 
  & $-33.3 + 48.5\ i$ & $-38.1 +38.5\ i$ \\
  1(d) & $-(\alpha_s/4\pi)C_F \ln^2 x_2$
  & $51.4 -51.2\ i$ & $48.1 -37.8\ i$ \\
  \hline
  1(c)+1(d) & & $18.1 -2.7\ i$ & $10.0 +0.7\ i$ \\
  \hline
  \hline
  1(e) & $-(\alpha_s/4\pi)(N_c/2) \ln^2 x_3$
   & $-114.3 +131.6\ i$ & $-102.2 +80.6\ i$ \\
  1(f) & $-(\alpha_s/4\pi)(N_c/2) \ln^2 x_3$
   & $128.2 -161.3\ i$ & $123.9 -117.6\ i$ \\
  \hline
  1(e)+1(f) & & $13.9 -29.7\ i$ & $21.7 -37.0\ i$ \\
  \hline
  \hline
  1(g) & $-(\alpha_s/4\pi)(N_c/2) \ln^2 (x_2 +{\bar x}_3)$ 
  & $7.6 -9.7\ i$ & $-$ \\
  1(h) & $-(\alpha_s/4\pi) [1/(2N_c)] \ln^2 x_2$
  & $-29.8 -33.4\ i$ & $-29.1 -34.5\ i$ \\
 \end{tabular}
\caption{Numerical effects from the threshold resummation 
on the different amplitudes (in the unit of $10^{-4}$ GeV).
Note that the associated Wilson coefficients have been set to unity.}
\label{tb:num_eff}
\end{table}

\begin{figure}[htbp]
\begin{center}
 \begin{picture}(375,390)(-100,-10)
   \put(-30,220){\includegraphics[height=5.5cm]{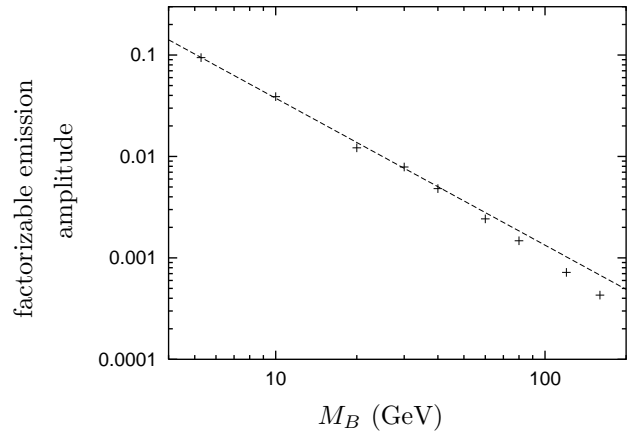}}
   \put(65,210){$M_B$ (GeV)}
   \put(75,193){(a)}
   \put(-50,250){\rotatebox{90}{factorizable emission}}
   \put(-35,280){\rotatebox{90}{amplitude}}
   \put(-30,20){\includegraphics[height=5.5cm]{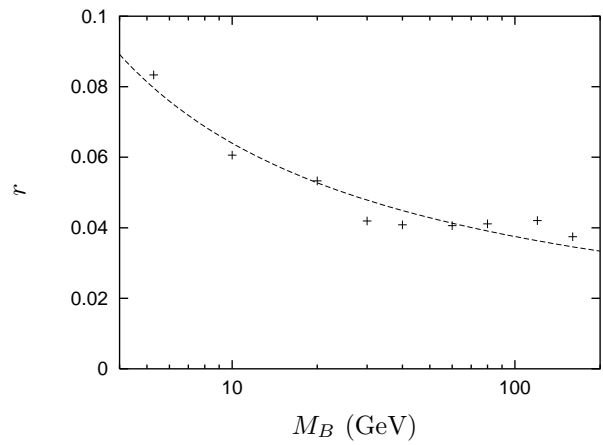}}
   \put(65,10){$M_B$ (GeV)}
   \put(75,-7){(b)}
   \put(-40,100){\rotatebox{90}{$r$}}
 \end{picture}
\end{center}
 \caption{(a) The factorizable emission amplitude as a function of 
$M_B$. (b) The ratio $r$ of the nonfactorizable emission amplitude
over the factorizable one as a function of $M_B$.}
 \label{fig:limit}
\end{figure}

Next we examine the factorization limit of the PQCD approach to two-body 
nonleptonic $B$ meson decays at large $M_B$. It is found that the 
factorizable emission amplitude decreases like $M_B^{-3/2}$
as displayed in Fig.~\ref{fig:limit}(a), if the $B$
meson decay constant $f_B$ scales like $f_B\propto M_B^{-1/2}$. This
power-law behavior is consistent with that obtained in \cite{BBNS,CZ}.
Figure~\ref{fig:limit}(b) exhibits the ratio of the magnitude of the 
leading-power nonfactorizable emission amplitude over the 
factorizable one as a function of 
$M_B$. The curve actually descends with $M_B$ despite of small 
oscillation. If parametrizing the ratio as
\begin{eqnarray}
r\equiv \frac{|\rm Nonfact.|}{\rm Fact.}\propto
\frac{1}{\ln^\alpha(M_B/\bar\Lambda)}\;,
\label{r}
\end{eqnarray}
the best fit to the curve gives the power $\alpha\sim 1.0$ for
$\bar\Lambda\sim 0.4$ GeV. We have confirmed this logarithmic decrease
up to $M_B=300$ GeV. It implies that the PQCD formalism approaches
the factorization assumption logarithmically. Note that the argument
in \cite{CKL} based on the power behavior of the integrand,
instead of the integral, just leads to the existence of the
factorization limit, providing no information on how fast to arrive
at this limit. We do not bother to examine the behavior of the 
annihilation amplitudes, since they are explicitly suppressed by 
$1/M_B$ compared to the factorizable emission.

Surprisingly, the behavior of the ratio $r$ with $M_B$ in PQCD is 
close to that in QCDF. However, the reasonings for achieving the same 
power counting are quite different. In the latter approach the 
factorizable contribution is assumed to be uncalculable in perturbation 
theory, and identified as being of $O(\alpha_s^0)$. The nonfactorizable 
contribution, being calculable, starts from $O(\alpha_s)$. Because of 
the soft cancellation at $x_3\sim O(\bar\Lambda/M_B)$, the
nonfactorizable emission amplitude is dominated by the contribution from
the region of $x_3\sim O(1)$. In this region there is no further power
suppression, and one has the ratio,
\begin{eqnarray}
r_{\rm QCDF}\sim \alpha_s(M_B)\propto 
\frac{1}{\ln(M_B/\Lambda_{\rm QCD})}\;.
\end{eqnarray}
In the former approach based on $k_T$ factorization theorem \cite{BS,LS},
both the factorizable and nonfactorizable contributions, being
calculable, start from $O(\alpha_s)$. However, the Sudakov factor
modifies the factorization formulas in the way that
a pair of nonfactorizable diagrams exhibits a stronger cancellation as
$M_B$ increases \cite{CKL}. It turns out that the ratio $r$ also
vanishes logarithmically as shown in Eq.~(\ref{r}).

\section{CONCLUSION}

In this letter we have analyzed the double logarithmic corrections to
two-body nonleptonic $B$ meson decays.
The results are the same for charmless and charmful modes at the leading
power. It has been observed that the double logarithms are universal for
the factorizable emission and annihilation topologies. Their effect
is crucial due to the presence of the potential linear end-point
singularities. For the nonfactorizable emission diagrams, the
logarithms carry a different (slightly larger) color factor. In the 
end-point region with a small parton momentum fraction, the
nonfactorizable contributions cancel by pair, such that the end-point 
singularities are at most logarithmic. The threshold resummation effect 
is then expected to be minor. For the nonfactorizable annihilation 
diagrams, the factorization formulas involve the overlap integrals
of three meson distribution amplitudes, and the singularities become
also logarithmic. Furthermore, the double logarithms are either
suppressed by phase space, or negligible with a tiny color factor.
That is, the double logarithms for the nonfactorizable amplitudes
depend on the topologies, and introduce only small
effects in the current leading-power PQCD formalism.

We have also examined the heavy-quark limit of the PQCD approach
to two-body nonleptonic $B$ meson decays numerically. The factorizable
emission amplitude shows the desired power behavior, proportional to
$M_B^{-3/2}$. PQCD reduces to the factorization assumption
\cite{BSW} as $M_B$ goes to infinity: the ratio of the nonfactorizable
emission contribution over the factorizable one diminishes 
logarithmically. The annihilation contribution
is explicitly power-suppressed. Our careful study justifies
that the PQCD approach has a valid factorization limit, and the criticism
in \cite{BBNS} is false. It is an interesting observation that the power
counting rule for the above ratio, though originating from different
reasonings, turns out to be identical in QCDF and in PQCD. With the
gauge invariance proved in \cite{L4} and the correct heavy-quark limit
demonstrated here, the PQCD approach stands as a solid theory.

\vskip 0.3cm
We thank M. Beneke, G. Buchalla, K. Hagiwara, and Z.T. Wei for useful 
discussions on the power counting rules.
This work was supported in part by the National Science Council of 
R.O.C. under Grant No. NSC-91-2112-M-001-053, by the National Center for
Theoretical Sciences of R.O.C., and by Theory Group of KEK, Japan, 
by Grand-in Aid for Special Project Research (Physics of CP Violation),
and by Grand-in Aid for Science Research from the Ministry of
Education, Culture, Sports, Science and Technology, Japan (No. 12004276).
The work of KU is supported by the Japan Society for Promotion of
Science under the Predoctoral Research Program. 


\begin{figure}[htbp]
\begin{center}
\begin{picture}(370,500)(-20,-260)
  \Line(20,140)(130,140)
  \Line(20,175)(72,175)\Line(78,175)(130,175)
  \Line(72,175)(60,215)
  \Line(78,175)(90,215)
  \Gluon(50,140)(50,175){3}{4}\Vertex(50,140){1.5}\Vertex(50,175){1.5}
  \put(23,180){$b$}\put(50,210){$s$}
  \put(0,150){{\large $B$}}
  \put(138,152){{\large $\pi$}}
  \put(70,220){{\large $K$}}
  \put(67,120){(a)}
  \Line(200,140)(310,140)
  \Line(200,175)(252,175) \Line(258,175)(310,175)
  \Line(252,175)(240,215)
  \Line(258,175)(270,215)
  \Gluon(280,175)(280,140){3}{4}\Vertex(280,140){1.5}\Vertex(280,175){1.5}
  \put(203,180){$b$}\put(230,210){$s$}
  \put(180,150){{\large $B$}}
  \put(318,152){{\large $\pi$}}
  \put(250,220){{\large $K$}}
  \put(247,120){(b)}
  \Line(27,70)(67,57)\Line(27,40)(67,53)
  \Line(67,57)(105,95)\Line(67,53)(105,15)
  \Line(100,55)(125,80)\Line(100,55)(125,30)
  \Gluon(100,55)(83,72){3}{3}\Vertex(100,55){1.5}\Vertex(83,72){1.5}
  \put(30,73){$b$}\put(95,95){$s$}
  \put(10,50){{\large $B$}}
  \put(117,92){{\large $K$}}
  \put(120,10){{\large $\pi$}}
  \put(60,0){(c)}
 \Line(207,70)(247,57)\Line(207,40)(247,53)
  \Line(247,57)(285,95)\Line(247,53)(285,15)
  \Line(280,55)(305,80)\Line(280,55)(305,30)
  \Gluon(263,38)(280,55){3}{3}\Vertex(280,55){1.5}\Vertex(263,38){1.5}
  \put(210,73){$b$}\put(275,95){$s$}
  \put(190,50){{\large $B$}}
  \put(297,92){{\large $K$}}
  \put(300,10){{\large $\pi$}}
  \put(240,0){(d)}
  \Line(20,-110)(130,-110)
  \Line(20,-75)(72,-75)\Line(78,-75)(130,-75)
  \Line(72,-75)(57,-35)
  \Line(78,-75)(93,-35)
  \GlueArc(115,-110)(70,135,180){2.5}{8}
  \Vertex(45,-110){1.5}\Vertex(66,-60){1.5}
  \put(23,-70){$b$}\put(47,-40){$s$}
  \put(0,-100){{\large $B$}}
  \put(138,-98){{\large $\pi$}}
  \put(70,-30){{\large $K$}}
  \put(67,-130){(e)}
  \Line(200,-110)(310,-110)
  \Line(200,-75)(252,-75)\Line(258,-75)(310,-75)
  \Line(252,-75)(237,-35)
  \Line(258,-75)(273,-35)
  \GlueArc(215,-110)(70,0,45){2.5}{8}
  \Vertex(285,-110){1.5}\Vertex(264,-60){1.5}
  \put(203,-70){$b$}\put(227,-40){$s$}
  \put(180,-100){{\large $B$}}
  \put(318,-98){{\large $\pi$}}
  \put(250,-30){{\large $K$}}
  \put(247,-130){(f)}
  \Line(28,-180)(68,-192)\Line(28,-210)(68,-198)
  \Line(68,-192)(105,-155)\Line(68,-198)(105,-235)
  \Line(100,-195)(125,-170)\Line(100,-195)(125,-220)
  \GlueArc(72,-205)(30,18,143){2.5}{8}
  \Vertex(100,-195){1.5}\Vertex(48,-186){1.5}
  \put(31,-177){$b$}\put(95,-156){$s$}
  \put(10,-200){{\large $B$}}
  \put(120,-155){{\large $K$}}
  \put(120,-240){{\large $\pi$}}
  \put(67,-250){(g)}
  \Line(208,-180)(248,-192)\Line(208,-210)(248,-198)
  \Line(248,-192)(285,-155)\Line(248,-198)(285,-235)
  \Line(280,-195)(305,-170)\Line(280,-195)(305,-220)
  \GlueArc(252,-185)(30,217,342){2.5}{8}
  \Vertex(280,-195){1.5}\Vertex(228,-204){1.5}
  \put(211,-177){$b$}\put(275,-156){$s$}
  \put(190,-200){{\large $B$}}
  \put(300,-155){{\large $K$}}
  \put(300,-240){{\large $\pi$}}
  \put(247,-250){(h)}
\end{picture}
\end{center}
\caption{Lowest-order diagrams which contribute to the $B\to K\pi$ 
decays, with (a) and (b) for the factorizable emission, 
(c) and (d) for factorizable annihilation,
(e) and (f) for nonfactorizable emission, and 
(g) and (h) for nonfactorizable annihilation.}
\label{fig:leading_Kpi}
\end{figure}
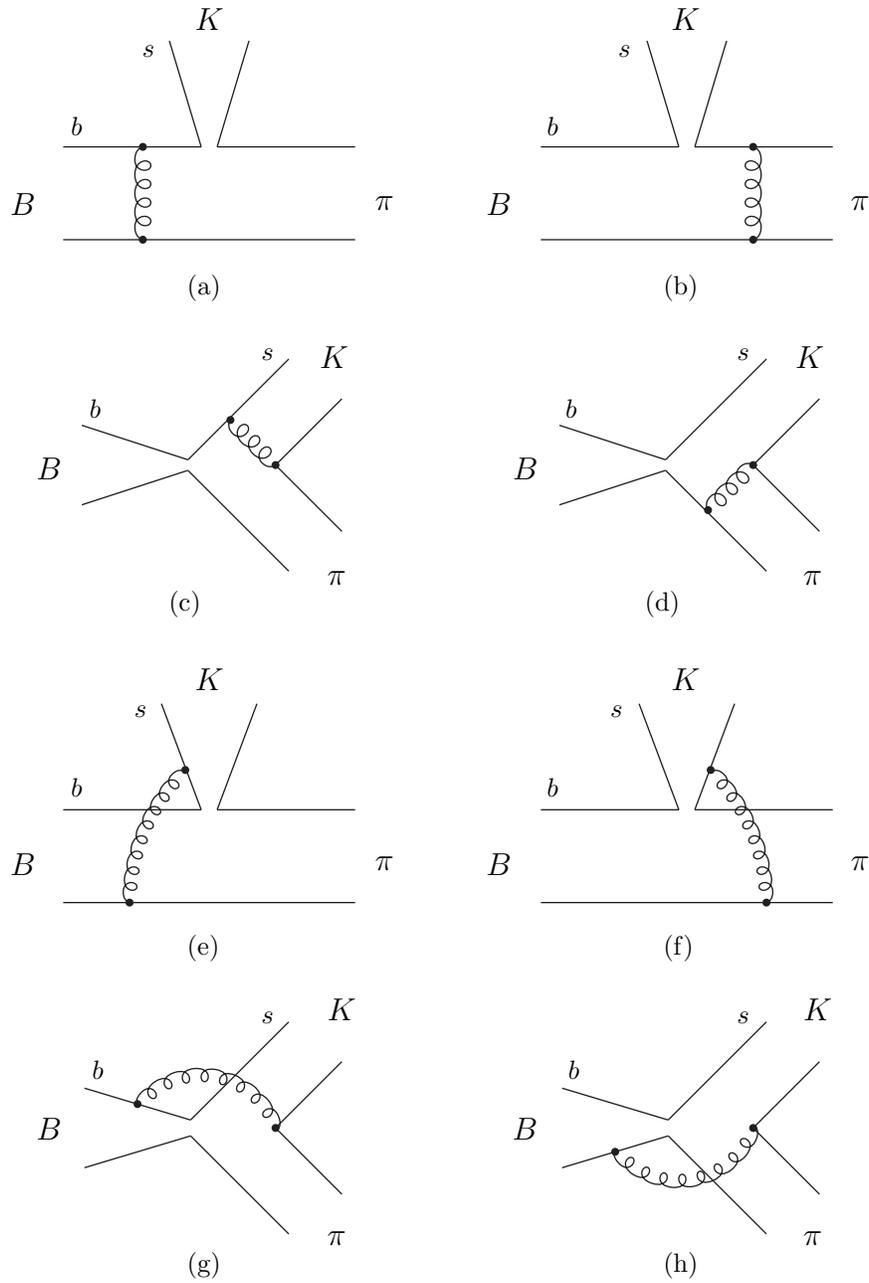

\begin{figure}[htbp]
\begin{center}
\begin{picture}(150,120)(0,120)
  \Line(20,140)(130,140)
  \Line(20,175)(72,175)\Line(78,175)(130,175)
  \Line(72,175)(60,215)
  \Line(78,175)(90,215)
  \Gluon(45,140)(45,175){3}{4}\Vertex(45,140){1.5}\Vertex(45,175){1.5}
  \put(23,180){$b$}\put(50,210){$s$}
  \put(0,150){{\large $B$}}
  \put(138,152){{\large $\pi$}}
  \put(70,220){{\large $K$}}
  \put(67,120){(a)}
 \GlueArc(75,180)(16,200,340){3}{5}
\end{picture}
\begin{picture}(150,120)(0,120)
  \Line(20,140)(130,140)
  \Line(20,175)(72,175)\Line(78,175)(130,175)
  \Line(72,175)(60,215)
  \Line(78,175)(90,215)
  \Gluon(40,140)(40,175){3}{4}\Vertex(40,140){1.5}\Vertex(40,175){1.5}
  \put(67,120){(b)}
 \GlueArc(56,175)(7,180,360){2}{4}
\end{picture}
\begin{picture}(150,120)(0,120)
  \Line(20,140)(130,140)
  \Line(20,175)(72,175)\Line(78,175)(130,175)
  \Line(72,175)(60,215)
  \Line(78,175)(90,215)
  \Gluon(45,140)(45,175){3}{4}\Vertex(45,140){1.5}\Vertex(45,175){1.5}
  \put(67,120){(c)}
 \GlueArc(45,170)(14,20,160){3}{4}
\end{picture}
\begin{picture}(150,120)(0,120)
  \Line(20,140)(130,140)
  \Line(20,175)(72,175)\Line(78,175)(130,175)
  \Line(72,175)(60,215)
  \Line(78,175)(90,215)
  \Gluon(40,140)(40,175){3}{5}\Vertex(40,140){1.5}\Vertex(40,175){1.5}
  \put(67,120){(d)}
 \GlueArc(40,175)(16,280,360){2.5}{3}
\end{picture}
\begin{picture}(150,120)(0,120)
  \Line(20,140)(130,140)
  \Line(20,175)(72,175)\Line(78,175)(130,175)
  \Line(72,175)(60,215)
  \Line(78,175)(90,215)
  \Gluon(40,140)(40,175){3}{5}\Vertex(40,140){1.5}\Vertex(40,175){1.5}
  \put(67,120){(e)}
 \GlueArc(78,175)(22,120,180){3}{3}
\end{picture}
\begin{picture}(150,120)(0,120)
  \Line(20,140)(130,140)
  \Line(20,175)(72,175)\Line(78,175)(130,175)
  \Line(72,175)(60,215)
  \Line(78,175)(90,215)
  \Gluon(40,140)(40,175){3}{5}\Vertex(40,140){1.5}\Vertex(40,175){1.5}
  \put(67,120){(f)}
 \GlueArc(76,175)(20,70,180){3}{5}
\end{picture}
\end{center}
 \caption{$O(\alpha_s)$ corrections to Fig.~\ref{fig:leading_Kpi}(a).}
 \label{fig:higher_a}
\end{figure}
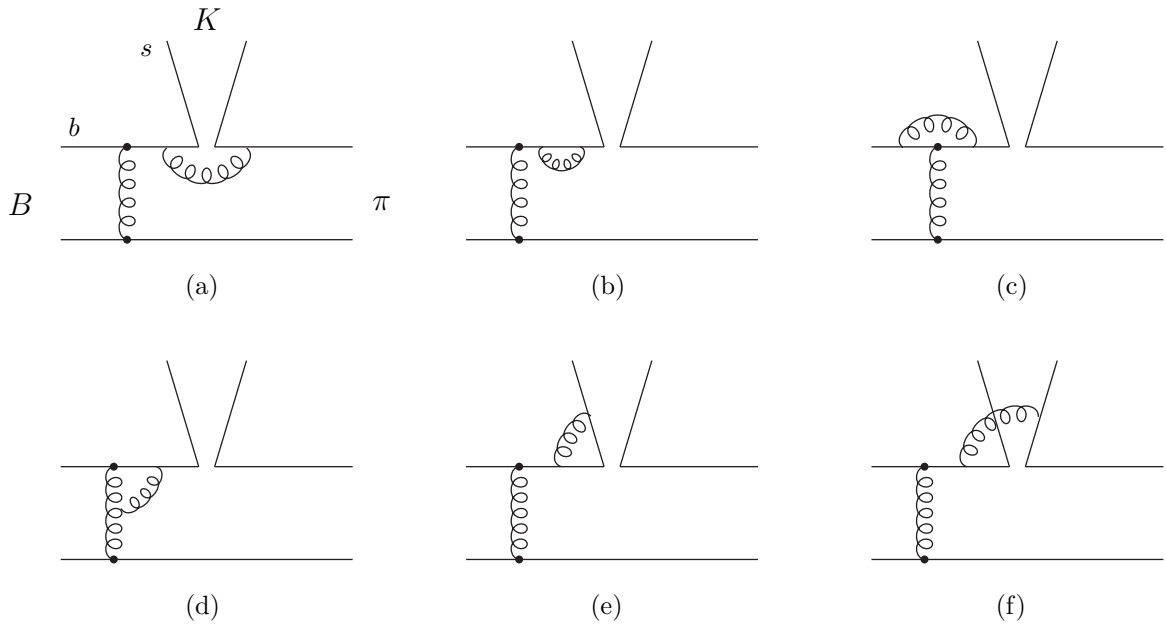

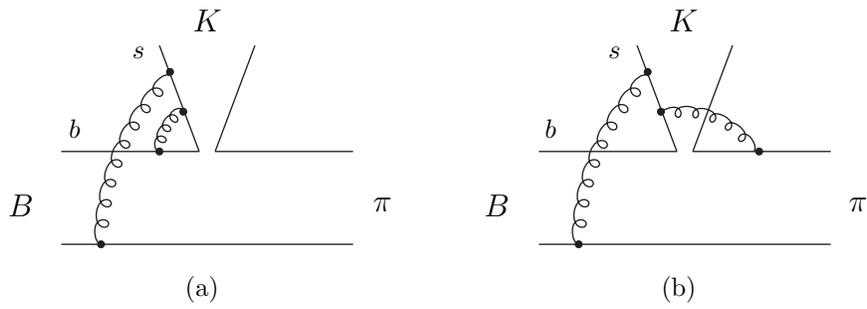
\begin{figure}[htbp]
\begin{center}
\begin{picture}(370,130)(-20,-130)
  \Line(20,-110)(130,-110)
  \Line(20,-75)(72,-75)\Line(78,-75)(130,-75)
  \Line(72,-75)(57,-35)
  \Line(78,-75)(93,-35)
  \GlueArc(125,-110)(90,135,180){2.5}{8}
  \GlueArc(75,-75)(18,120,180){2}{3}
  \Vertex(57,-75){1.5}\Vertex(66,-60){1.5}
  \Vertex(35,-110){1.5}\Vertex(61,-45){1.5}
  \put(23,-70){$b$}\put(47,-40){$s$}
  \put(0,-100){{\large $B$}}
  \put(138,-98){{\large $\pi$}}
  \put(70,-30){{\large $K$}}
  \put(67,-130){(a)}
  \Line(200,-110)(310,-110)
  \Line(200,-75)(252,-75)\Line(258,-75)(310,-75)
  \Line(252,-75)(237,-35)
  \Line(258,-75)(273,-35)
  \GlueArc(305,-110)(90,135,180){2.5}{8}
  \GlueArc(255,-90)(30,30,108){2}{5}
  \Vertex(283,-75){1.5}\Vertex(246,-60){1.5}
  \Vertex(215,-110){1.5}\Vertex(241,-45){1.5}
  \put(203,-70){$b$}\put(227,-40){$s$}
  \put(180,-100){{\large $B$}}
  \put(318,-98){{\large $\pi$}}
  \put(250,-30){{\large $K$}}
  \put(247,-130){(b)}
\end{picture}
\end{center}
 \caption{$O(\alpha_s)$ corrections to Fig.~\ref{fig:leading_Kpi}(e) 
 which give rise to the double logarithms.}
 \label{fig:nonfact_cort}
\end{figure}

\end{document}